\documentclass{elsart4-1}

\usepackage{graphicx}

\usepackage[english,francais]{babel}

\newtheorem{e-proposition}[theorem]{Proposition}

\newtheorem{e-definition}[theorem]{Definition\rm}

\renewcommand{\theequation}{\arabic{equation}}
\def\spose#1{\hbox to 0pt{#1\hss}}
\def\lta{\mathrel{\spose{\lower 3pt\hbox{$\mathchar"218$}}
        \raise 2.0pt\hbox{$\mathchar"13C$}}}
\def\gta{\mathrel{\spose{\lower 3pt\hbox{$\mathchar"218$}}
        \raise 2.0pt\hbox{$\mathchar"13E$}}}

\def\msol{\rm M_\odot}

\def\msol{\rm M_\odot}

\def\og{\leavevmode\raise.3ex\hbox{$\scriptscriptstyle\langle\!\langle$~}}
\def\fg{\leavevmode\raise.3ex\hbox{~$\!\scriptscriptstyle\,\rangle\!\rangle$}}

\begin{document}

\centerline{Physics or Astrophysics/Header}
\begin{frontmatter}


\selectlanguage{english}
\title{Physics of accretion flows around compact objects}


\selectlanguage{english}
\author[authorlabel1]{Jean-Pierre Lasota},
\ead{lasota@iap.fr}

\address[authorlabel1]{Institut d'Astrophysique de Paris, UMR 7095 CNRS,
Universit\'e Pierre \& Marie Curie, 98bis Bd Arago, 75014 Paris,
France}


\medskip
\begin{center}
{\small Received *****; accepted after revision +++++}
\end{center}

\begin{abstract}
Several physical and astrophysical problems related to accretion
onto black holes and neutron stars are shortly reviewed. I discuss
the observed differences between these two types of compact objects
in quiescent Soft X-ray Transients. Then I review the status of
various non-standard objects suggested as an alternative to
black-holes. Finally I present new results and suggestion about the
nature of the jet activity in Active Galactic Nuclei. {\it To cite
this article: J.-P. Lasota, C. R. Physique 6 (2005).}

\vskip 0.5\baselineskip

\selectlanguage{francais} \noindent{\bf R\'esum\'e} \vskip
0.5\baselineskip \noindent {\bf Physique des flots d'accr\'etion
autour de objets compacts}  L'article contient une br\`{e}ve revue
de quelques probl\`{e}mes li\'{e}s \`{a} l'accr\'{e}tion sur les
\'{e}toiles \`{a} neutrons et les trous noirs. Je discute les
diff\'erences entre ces deux types d'objets compacts quand ils sont
observ\'es dans les Sources X Transitoires quiescentes. Ensuite,
j'examine l'int\'er\^et astrophysique, mais aussi fondamental, des
divers objets non standard propos\'es comme alternatives aux trous
noirs. La parties finale de l'article contient une pr\'esentation de
certains r\'esultats r\'ecents concernant la nature des jets
\'emanants des Noyaux Actifs de Galaxies. {\it Pour citer cet
article~: Jean-Pierre Lasota, C. R. Physique 6 (2005).}

\vskip 0.5
\baselineskip \noindent{\small{\it Mots-cl\'es~:}
trous noirs; Relativit\'e G\'en\'erale; accr\'etion; jets relativistes}}
\end{abstract}
\end{frontmatter}


\selectlanguage{english}
\section{Introduction}
\label{intro}

The physics of accretion onto compact objects is interesting for at
least two reasons. First, the compact bodies themselves are objects
fascinating by their extreme properties: supernuclear densities and
very strong magnetic fields are the apanage of neutron stars,
whereas black holes are a marvel of pure relativistic gravitation.
The signatures of neutron stars are usually unmistakeable: very
regular pulses of electrodynamic radiation or X-ray bursts due to
thermonuclear explosions at their surface. In other cases the
presence of neutron stars is deduced from properties analogous to
systems in which their presence is well established. Some of the
reputed neutron stars could be quark stars. Nobody speaks, however,
of ``candidate neutron stars". The status of black holes is not the
same. Although they are a very conservative prediction of Einstein'
theory of gravitation, and the calculation showing that a collapsing
cloud of dust forms a black hole appeared in 1939~\cite{os}, they
have a rather louche reputation. One often (though less often than a
few years ago) speaks of ``black-hole candidates". This is not the
place to analyze the reasons for the status of this fascinating
object. It is certainly due to the fact that they are fascinating,
and they are fascinating because they are related to so many
fundamental questions of physics. I will come to his later in the
article.

From the point view of an astrophysicist, however, they are in a
sense rather boring. Their only properties are mass and angular
momentum. Maybe a little, but only a little, charge. As an
astrophysicist I don't care that when a star collapses to form a
black hole, entropy apparently increases by tens of orders of
magnitude. I cannot measure it. I can measure its mass and estimate
its angular momentum, but even if I can use these quantities to
calculate a quantity called ``black hole entropy" it has no physical
meaning because there is no way I can measure it. Black holes have
been identified in many binary systems and in the centers of many
galaxies. They are supposed to be featureless (except for their mass
and angular momentum) and in not a single case have they failed to
live up to this reputation. More precisely, not a single compact
object more massive than 3$\msol$ has shown any feature that would
allow us to attribute to it a property other than mass and rotation.
The ultimate test of their properties will be obtained by measuring
gravitational waves emitted during black-hole mergers or black-hole
ringing when excited by an orbiting compact body\cite{sh}.

This said, it is legitimate and interesting to investigate the
evidence for the of  existence black holes, i.e. whether
observations can exclude the existence (or rather presence) of other
less orthodox and more exotic bodies. An article I co-authored a few
years ago~\cite{akl02} was devoted to this problem. In it we
critically discussed some evidence and arrived at the conclusion
that the ultimate test cannot be obtained by electromagnetic
observations. Unfortunately our article was understood by some
people as doubting the existence of black-holes. This article will
try in part to dissipate that impression. However, I am not going to
review here the whole problem of proving the existence of black
holes. For this I refer the reader to the excellent review by Ramesh
Narayan\cite{narayan}.

\section{ADAFs and observations}

ADAFs (Advection Dominated Accretion Flows - a name I devised at the
1995 Kyoto conference on the \textsl{Physics of accretion disks :
advection, radiation and magnetic fields}) are accretion flows in
which most of the thermal energy is not radiated but advected onto
(in the case of a star) or through (in the case of a black hole) the
surface of the accreting body. (Strictly speaking in some types of
such flows advection of thermal energy into the central black hole
is negligible \cite{bptoy}, \cite{chen1}). The idea that such flows
have some relevance to astrophysics had been around for some
time\cite{rees82}\cite{ichi}\cite{slim}\cite{nap}. In 1994--95 it
was formalized (see ref. \cite{chenetal} and references therein) but
only the brilliant and successful application of ADAFs to various
astrophysical systems by Narayan and collaborators (since I had the
privilege of being one one of them I am obviously not impartial)
showed how useful and powerful the concept is (see ref.\cite{menouj}
and references therein). The enthusiasm was not universal and the
idea has been challenged many times (also by the ADAF authors
themselves) but the (slightly bruised) ADAF ``paradigm" is still
around and it is unlikely that it will soon disappear (see e.g. ref.
\cite{narayan2}).

In binary systems the ADAF forms only the inner part of the
accretion flow the outer, part being a radiatively efficient
(geometrically) thin disc. In galactic nuclei pure ADAF models have
been proposed as in the case of Sgr~A$^*$ (\cite{goldstone} and
references therein). I will limit here myself to the case of binary
systems

\subsection{Accreting black holes are fainter than accreting neutron stars}

\begin{figure}
\center
\resizebox{0.8\hsize}{!}{\includegraphics[angle=0]{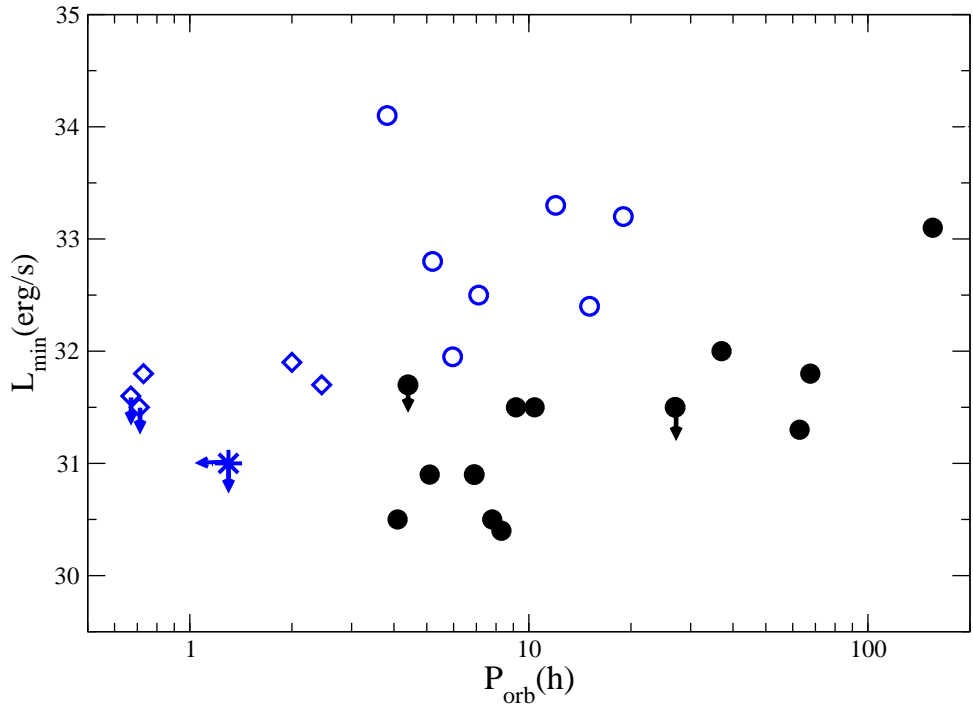}}
\caption{Quiescent luminosities of black holes (filled circles) and
neutron star
 (open circles and open diamonds) soft X-ray transients. Diamonds
 corresponds to accreting millisecond pulsars. The star represent
 the system 1H~1905+000 whose orbital period is unknown (see the
 text). This figure shows that in quiescent
 transient LMXBs, \textsl{at a given orbital
 period}, neutron stars are brighter than black holes.
 Data for black holes from Garcia (private communication),
 for neutron stars from ref.  \cite{Campana}}
\label{fig:lumin}
\end{figure}
If one accepts the ADAF picture for both accreting black-holes and
neutron stars one consequence is immediate: neutron stars should be
brighter because all the thermal energy that was not radiated in the
accretion flow will have to be emitted from the stellar surface.
Guided by this idea Narayan, Garcia, \& McClintock\cite{ngm}
compared the outburst amplitudes of Black-Hole Soft X-ray Transients
(BH SXTs) and Neutrons Star Soft X-ray Transients (NS SXTs) as a
function of their maximum luminosities. They found systematically
higher amplitudes in BH SXTs implying that in quiescence (when ADAFs
are supposed to be present) they are fainter than NS SXTs. Lasota \&
Hameury~\cite{somelike} pointed out that the ADAF model does not
state that accreting black holes are \textsl{always} fainter than
accreting neutron stars, but that this the case when both type of
objects accrete at the same rate. In order to test the ADAF
hypothesis they suggested plotting the quiescent luminosity as a
function of the orbital period. This idea was later developed in
ref.\cite{menouetal}. Figure \ref{fig:lumin} shows the quiescent
luminosity vs orbital period for 13 NS SXTs and 13 BH SXTs. I will
discuss some aspects of this figure below but the separation between
the two classes of systems is clear and neat: \textsl{at each
orbital period} neutron-star systems are brighter than their
black-hole counterparts. Doubtless it is a strong argument in favour
of ADAFs and black holes. There has been a debate about where the
neutron-star quiescent luminosity comes from, what part of it, if
any, is due to accretion (see \cite{rutledge} and references
therein; \cite{menouetal}) but in BH SXTs the quiescent luminosity
is certainly due to accretion and that it is unlikely that quiescent
discs around neutron stars are not truncated~\cite{dubus01} (or
truncated but not leaky).

One should stress that the idea of plotting the luminosity as a
function of period is \textsl{not} based on the assumption that BH
SXTs and NS SXTs have the same transfer rates at a given orbital
period. It is based on the assumption that the truncation radius
where the transition from disc to ADAF occurs, is roughly a constant
fraction of the circularization
radius~\cite{menouetal}~\cite{faint}~\cite{lasota00}. Another
assumption is that the truncated disc is a non-equilibrium disc as
described by the disc instability model (see \cite{NAR} for a
review). These two assumptions (the second, although often not
understood, should be noncontroversial, the first, if one accepts
the ADAF model, rather obvious) make sure that quiescent BH and NS
SXTs with similar orbital period accrete matter at a similar rate.

Recently Jonker et al.~\cite{jonker} found that the neutron-star
soft X–ray transient 1H~1905+000 could be the spoilsport
long-awaited by the ADAF-basher crowd. Its quiescent X-ray
luminosity is at most $1.8 \times 10^{31}$ erg s$^{-1}$, but the
upper limit on the 0.5 -- 10 keV luminosity of this source,
undetected by \textsl{Chandra}, could be as low as $1.0\times
10^{31}$ erg s$^{-1}$. Jonker et al. assert that the luminosity of
this neutron-star binary is so low that it is similar to the lowest
luminosities derived for black hole SXTs in quiescence and that it
``challenges the hypothesis presented in the literature that black
hole SXTs in quiescence have lower luminosities than neutron star
SXTs as a result of the presence of a black hole event horizon."
However, looking at Fig. \ref{fig:lumin}, I see no reason to panic.
The orbital period of 1H~1905+000 is unknown but it is certainly
very short~\cite{jonker}. I have therefore tentatively assumed a
period of $1.3$ h, but even a longer period ($\lta 3$ hr) would not
contradict the claim that black-hole systems are fainter than those
harbouring neutron stars.

Even an actual quiescent X-ray luminosity of 1H~1905+000 much lower
than the \textsl{Chandra} upper limit would not  necessarily be a
problem for the black-hole faintness ``paradigm"; it depends on what
kind of system 1H~1905+000 is. The faintness of the secondary
implies a short compact binary containing either a hot brown-dwarf
companion, similar to e.g. SAX~1808-365~\cite{sax} or an
ultra-compact X-ray binary (UCXB), in which case the neutron star
companion would be a low-mass helium or carbon-oxygen white
dwarf~\cite{ucb}. When transient (very short-period systems are
rather persistent~\cite{deloy}, such compact binaries exhibit short
($\gta$10 -- $\gta$100 days), exponentially decaying outbursts as
expected from small, X-ray irradiated accretion discs
\cite{dubus01}. In all these very compact transient systems the
neutron star is a millisecond pulsar (MSP). Both their outburst
(usually $\sim$ few $\%$ of the Eddington luminosity) and quiescent
X-ray luminosities ($< 10^{32}$~erg s$^{-1}$) are lower than those
observed in longer period SXTs~\cite{Campana}. This is very similar
to 1H~1905+000 whose outburst luminosity was $\sim 4\times 10^{36}$
erg s$^{-1}$. However, the outburst behaviour of this system is
totally different from that observed in other short-period binaries
and UCXBs. Instead of short outbursts 1H~1905+000 exhibited one
$\gta$ 10 year long outburst that ended the late 1980s or early
1990s. Since then it has been quiet.

It is not clear why 1H~1905+000 is so different. During 11 years,
say, it had accreted $\sim 8 \times 10^{24}$ g. This is a lot, but a
hydrogen-dominated accretion disc can contain as much as
\begin{equation}
M_{\rm D, max}\approx 2.5 \times 10^{24} \left(\frac{\alpha_{\rm
cold}}{0.01}\right)^{-0.83} \left( \frac{M_{\rm ns}} {\rm 1.4
M_\odot} \right)^{-0.38} \left( \frac{P_{\rm orb}}{1.3\rm h}
\right)^{2.09}\ {\rm g}, \label{eq:diskmass}
\end{equation}
where $\alpha_{\rm cold}$ is the cold-disc viscosity parameter. For
the disc radius I used
\begin{equation}
\frac{R_d}{a}=\frac{0.60}{1+q},
\label{rd}
\end{equation}
where $a$ is the binary separation. The critical density of a cold
helium disc being $\sim 50$ times higher a UCXB disc would contain
even more mass (however, standard-disc formulae apply only to
mass-ratios $q\gta 0.02$, see below).

The maximum outburst luminosity for hydrogen-dominated disc, a 1.4
$\msol$ neutron star and $\alpha_{\rm hot}= 0.2$ can be estimated as
\begin{equation}
L_{\rm max} \simeq 1.8 \times 10^{36} \left( \frac{P_{\rm orb}}{1
\rm h}\right)^{2.09}\rm erg\ s^{-1}, \label{lmax}
\end{equation}
where I crudely re-scaled the formula from \cite{NAR} to take into
account disc irradiation. The maximum luminosity for a helium or
carbon oxygen disc (when it exists) would be a factor $\sim 2$
higher. Therefore 1H~1905+000 could in principle be a ``normal",
short-duration X-ray transient source, but it isn't. Maybe its long
"outburst" was due to irradiations of the secondary. If its period
is $\sim 20$ min it could be marginally stable with respect to the
thermal-viscous instability in an irradiated helium (or
carbon-oxygen) accretion disc.

There is, however, a more fundamental problem. The form of
mass-transfer in systems with such very low mass ratios has not been
studied and only some general properties of such systems can be
conjectured~(Dubus, private communication). For values of $q\lta
0.02$ the circularization radius becomes greater than the estimates
of the outer radius given by \cite{Paczynski77} and
\cite{pappringle77}. Most probably matter streaming in from the
companion circularizes onto unstable orbits. At $q\approx 0.02$,
matter is added at $R_{\rm circ}$ onto orbits that can become
eccentric due to the 3:1 resonance. At $q\approx 0.005$ the
circularization radius approaches the 2:1 Lindblad resonance. This
might efficiently prevent mass being transferred onto the compact
object.

An equivalent system with a black-hole instead of a neutron star
would have a minuscule mass ratio $< 0.01$ ($M_{\rm bh}> 4 \msol$).
It might not be a coincidence that there are no observed black-hole
counterparts of neutron-star X-ray binary systems at orbital periods
shorter than 2 hours. Such systems might exist~\cite{ljova} but they
are not your normal LMXBs.

The 1H 1905+000 challenge is very likely a red herring and a
counterexample to the black-hole faintness ``paradigm" has yet to be
found.

\section{Demography: the accreting bodies}
\label{demo}

We wish to know what is accreting because knowing that the object is
a black hole would be of fundamental interest, but also the physics
of accretion depends on what the matter falls onto. Some aspects of
the physics of accretion depend on the nature of the accreting
compact objects. No matter can accumulate at the surface of a black
hole which prohibits thermonuclear bursts. No boundary layer can
exist between an accretion disc and the (null) black-hole surface
where the accreted matter must plunge in radially at the speed of
light. Since a black hole is strictly axisymmetric no periodic
signal can be emitted by a (stationary) black hole. Black holes have
no magnetic fields so there is no magnetic disruption of the
accretion flow (external currents, however, can create magnetic
fields anchored on black-holes -- such fields can influence
accretion flows; see e.g. \cite{li}). On the other hand rotating
black-holes are surrounded by a region of absolute no-rest - the
ergosphere - which opens possibilities denied to celestial bodies
with material surfaces, such as the Blandford-Znajek
process\cite{bz77}. In this section I will mainly (but not only)
deal with compact bodies in compact binaries.

\begin{figure}
\center
 \resizebox{0.8\hsize}{!}{\includegraphics[angle=-90]{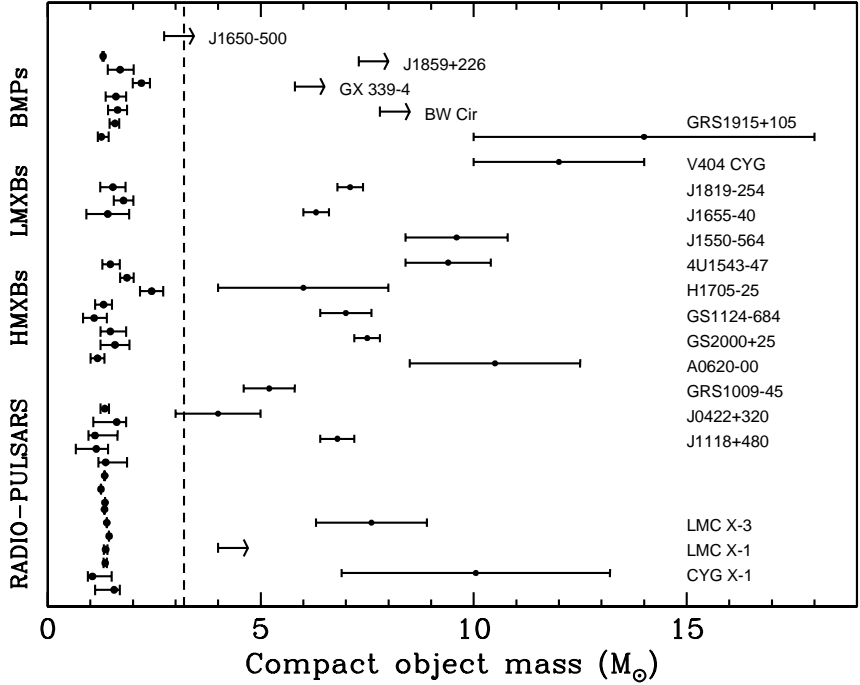}}
 \caption{Masses of neutron stars and black holes. From Casares~\cite{casares}}
 \label{fig:masses}
\end{figure}

\subsection{Neutron and quark stars}

Compact objects in close binary systems show mass segregation (see
Fig. \ref{fig:masses}). They are clearly divided into two mass
ranges: $ M < 3 \msol$ and $ M > 3 \msol$. All members of the lower
mass set are confirmed neutron stars, i.e. they are either (radio or
X-ray) pulsars or X-ray bursters\footnote{Some of them could be
quark (``strange") stars.} The member of the higher mass set are
certainly not neutron stars since the maximum mass of these
celestial bodies satisfies the inequality\cite{haensel}
\begin{equation}
M_{\rm max}\le M^{\rm CL}_{\rm max}= 3.0\left( {5\times 10^{14}~{\rm
g~cm^{-3}}\over \rho_{\rm u}} \right)^{{1\over 2}}~{\rm M}_\odot~.
\label{eq:CLMmax}
\end{equation}
where ``CL" stands for ``causality limit"; $\rho_{\rm u} \lta
2\rho_0$ ($\rho_0=2.7\times 10^{14}~\rm g~cm^{-3}$ is the normal
nuclear density) is the fiducial density above which equation of
state (EOS) of (super)nuclear matter EOS is unknown, i.e. there
exist a whole bunch of EOS that could describe matter at these
supernuclear densities. If $\rho_{\rm u}\lta 2\rho_0$ the
contribution of the outer layers ($\rho < \rho_{\rm u}$) of the
neutron star to the maximum mass is negligible (for this and the
following see the excellent lectures by Haensel\cite{haensel}). The
maximum mass of neutron stars with causal ($c_{\rm s}\leq c$, where
$c_{\rm s}$ is the adiabatic speed of sound) EOSs is only slightly
increased by rotation: $M^{\rm rot}_{\rm max} \simeq 1.18 M^{\rm
stat}_{\rm max}$\cite{lha}. The upper bound  $M^{\rm CL}_{\rm max}$
is increased by rotation up to $\sim 30\%$.

$M^{\rm CL}_{\rm max}$ can be considered to be the maximum mass of a
star with surface density $\rho_s=\rho_u$ and a pure causality-limit
EOS ($c_{\rm s}=c$, ). In other words the maximum mass of $M_{\rm
max}\sim (\rho_s)^{-1/2}$. Quark stars cannot be very massive
because their surface density is still in the nuclear regime but by
lowering the allowed surface density, i.e. $\rho_{\rm u}$ one can
easily obtain maximum masses in the black-hole ``range".

In any case we know that hypothetical objects more massive than
3$\msol$ cannot be made of normal matter, or matter in a normal
state (this include also quark matter in various states). Therefore
normal matter accreting onto the surface of such bodies might as
well be undergoing a phase transition transmuting itself to whatever
forms the accreting object. Therefore calculating models of
thermonuclear explosions at the surface of ``stars" more massive
than 3$\msol$\cite{rn03}\cite{nh03} is a rather pointless. There may
be no nuclei to burn\cite{akl02}. But of course the absence of X-ray
bursts from systems believed to host black hole is a strong argument
in favour of their real presence there.

\subsection{Exotic interlopers}

There exist several exotic alternatives to black holes. The oldest
are boson stars, which were not invented as black-hole competitors
(e.g. \cite{remobona}). Other compact exotic bodies were explicitly
introduced as black-hole challengers.

\subsubsection{Q-stars}

This was the story of Q-stars~\cite{qstars}. They are objects made
of a hypothetical state of matter which is a macroscopic self-bound
superdense scalar-field condensate with a well defined electric and
baryonic charge. They generated some astrophysical interest because
the maximum mass of Q-stars could be as large as $\sim 10\msol$.
However, in fact these high mass value were obtained by allowing
sufficiently low values of the fiducial density
$\rho_u$\cite{haensel}. Of course we don't know what it could be in
reality. No experimental evidence of the existence of Q-matter has
been been found until now but the same is true of cosmological dark
matter.

\subsubsection{Boson stars}

The typical mass of a star made of non-interacting bosons is $\sim
m_{\rm Pl}^2/m_b$, where $m_{Pl}=(\hbar c/G)^{1/2}$ is the Planck
mass and $m_b$ is the boson mass, but the maximum mass of a
self-interacting boson star is $\sim \lambda^{1/2}m_{\rm
Pl}^3/m_b^2$, where $\lambda \sim 1$ is the scalar field
self-coupling. This is similar to the mass of a barion star ($\sim$
Chandrasekhar mass) $\sim m_{\rm Pl}^3/m_p^2 = 1.9\ {\rm M}_\odot$,
where $m_p$ is the proton mass, so that for $m_b\sim m_p$ a boson
star would have a mass comparable to that of a neutron star. Since
by construction a boson star is a macroscopic quantum state
``supported" against gravity by the uncertainty principle its radius
$\sim 1/m_b$ is (for relativistic bosons) comparable to the radius
of neutron star. Although for the sake of completeness Yuan, Narayan
\& Rees~\cite{ynr} calculated models of accreting boson stars in
close binaries (and came to the conclusion that they would have Type
I X-ray bursts) the presence of these exotic bodies in such systems
is rather doubtful because it would require a fermion star evolving
into a boson star. I have no doubt that the imagination (and skills)
of some of my colleagues would find an easy way around this
objection but if one wants to consider a boson star as a serious
substitution for a black hole, its place is the Galactic Center.
Indeed, for a boson mass $m_b \sim$ 1 MeV the mass of a boson
``star" would be $\sim 10^6 {\rm M}_\odot$ and its radius a bit
larger than that of the Sun (see e.g. \cite{schliddle,toretal00}).
Such an object, and except for black hole only such an object, would
fit nicely into the constraints on the nature of the observed dark
mass in the Galactic Center \cite{schodel03}. The ultimate test of
the black-hole's presence will be brought by the gravitational-wave
observatory $LISA$\cite{sh,3h}.

\subsubsection{Dark energy stars aka gravastars}

Dark-energy stars (DES) (also called gravastars) have been proposed
as an alternative to black holes~\cite{chapline}~\cite{gravastars}.
In these objects the event horizon of a black hole is replaced by a
quantum phase transition of the vacuum of space-time analogous to
the liquid-vapor critical point of a Bose fluid. Since outside such
objects General Relativity is supposed to be valid, they are
described by the Schwarzschild solution down to a distance
$\epsilon$
\begin{equation}
\epsilon\sim \left(\frac{M_{\rm Pl}}{M}\right)^{2/3}\sim
10^{-25}\left(\frac{\msol}{M}\right)^{2/3}
\label{epsilon}
\end{equation}
from the Schwarzschild radius. The inner negative-pressure
gravitational vacuum condensate is protected by a very thin shell,
which effectively forms the DES surface\cite{gravastars}. Since this
surface is at $R_{\rm g*}= R_{\rm S}(1 + \epsilon)$ gravastar sizes
are not restricted by the
Buchdahl-Bondi~\cite{buchdahl1}~\cite{bondi} bound\footnote{This
bound was first noticed by Karl Schwarszchild\cite{ks}.} ($R_*
> 9/8R_{\rm S}$, or the redshift $z > 0.33$ - for comparison the
maximum redshift of a neutron star is 0.851~\cite{maxz}) their
ridiculously large redshifts make them apparently indistinguishable
from black holes, hence their potential astrophysical
interest\cite{akl02}. Such compactness is achieved at a price: to
avoid the presence of naked singularities gravastars must have
anisotropic pressures and a very peculiar equation of
state~\cite{anigrava}. Of course if one accepts negative central
pressures and violation of the dominant energy condition (which
requires $\rho \geq 0; \ |{p}| \leq \rho$) this cannot be the reason
to dismiss dark energy stars (incidentally, but not accidentally,
``normal" boson stars are subject to anisotropic stress and cannot
be described by an equation of state \cite{remobona}).

The problem with dark energy stars is indeed more fundamental: their
existence does not emerge from a (new) theory. They are constructed
by analogy with superfluidity, liquid helium, optical fibers etc.
and rather belong to the category of ``Analog Gravity"
models~\cite{analog} and should be treated as such. Although DES are
advertised as panacea for the inconsistencies between quantum
mechanics and General Relativity, Einstein's theory of gravitation
deserves better than to be replaced (if and when necessary) just by
an analogy. Especially that the analogy is far from perfect. In the
latest installment of the DES saga~\cite{godel2} it has been
proposed that the answer to the ``long-standing puzzle of
astrophysics; namely, how (...) during the gravitational collapse of
a massive stellar core the baryon number of the core disappears in
$\sim 10^{-5}$ sec" be that the nucleons undergoing gravitational
collapse disappear, being converted to vacuum energy when according
to General Relativity a trapped surface forms. If true, this would
be in violent contradiction with the equivalence principle (to an
observer in a free falling frame everything appears normal when
crossing the horizon) and require something better than just the
affirmation that (in some reference frames) space-time behaves like
a superfluid. Incidentally the ``disappearance" of baryons during
collapse has never been an astrophysical puzzle.

DES try to find their place in the dark, multidimensional landscape
where tachyons chase phantoms\footnote{phantoms are states with
negative free kinetic energy} in their quest to couple to various
types of (presumably) supersymmetric matter, so they do not deserve
the indifference they were met with. They are no more eccentric than
most ``theories" appearing everyday on arXiv.org and they are more
interesting than most. Their astrophysical interest, however, is
rather limited. No solution for a rotating gravastar has been found
(the G\"{o}del-like metric of \cite{godel1} is not such a solution
as it is not matched to an external vacuum solution). Since DES
properties are only vaguely defined, the suggested astrophysical
``tests" cannot be taken seriously. Of course the same can be said
about many models elaborated by astrophysicists but at least these
do not claim to result from the revision of fundamental laws of
physics.

The main motivation behind DES and similar enterprises is, however,
not astrophysical. Some physicists are depressed by the presence of
the singularity hidden behind the event horizon and some are unhappy
both with the singularity and the event horizon. This motivates them
to find an alternative to black holes. At a more fundamental level
the worry is the apparent incompatibility between General Relativity
and quantum mechanics. The unitary character of quantum mechanics
does not agree well with the presence of event horizons and
especially black-hole evaporation. A fashionable response to this is
that (super)string theory has the answer to these problems. It
seems, however, that for the moment this is only a hope and that
``(understanding) how string theory prevents quantum information
from being destroyed by black holes" and ``(understanding) when and
how string theory resolves spacetime singularities" are still
``remaining problems" of string theory~\cite{schwarz}. There are
also attempts to suitably generalize quantum mechanics so that black
hole evaporation would not be in conflict with its
principles~\cite{hartle}.

Obviously the source of all these difficulties is the absence of a
quantum theory of gravitation. Both Einstein's theory of gravitation
and quantum mechanics are extremely well tested experimentally in
their respective domains of application. One does not need General
Relativity to describe an atom and quantum mechanics is not good in
describing planetary orbits. In fact there is not a single observed
phenomenon or experimental fact that requires a quantum gravity
explanation. Hawking radiation, although generally treated as fact,
has never been obviously observed. Gravitational waves produced by
inflation have yet to be detected.

It is possible that we will have to live with two distinct theories
describing the Universe at different scales and that their
unification is physically meaningless\cite{dyson}\cite{graviton}. If
this is true, black holes are purely classical objects.

\section{Accretion, jets and spin}
\label{jets}

Rotating accretion flows often show evidence of the presence of more
or less collimated outflows. One speaks of \textsl{jets} when the
observed outflow is very collimated but no precise definition of jet
exists; sometimes the jet could be just the collimated outflow's
core that managed to bore through the surrounding medium or simply
the central part of a more extended outflow (see e.g.
\cite{krasnop}). Accreting compact objects produce relativistic
jets, i.e. well-collimated outflows with bulk motion corresponding
to $\Gamma= \sqrt{1/1 - (v_{\rm bulk}/c)^2} > 1$. Contrary to naive
expectations the $\Gamma$ of black-hole jets does not seem to be
larger than the $\Gamma$ of ejecta from neutron stars; there is even
evidence to the contrary, but one should keep in mind that
determination of flow's relativistic factors are rather deductions
than direct measurements.
\begin{figure}
\center
 \resizebox{0.8\hsize}{!}{\includegraphics[angle=0]{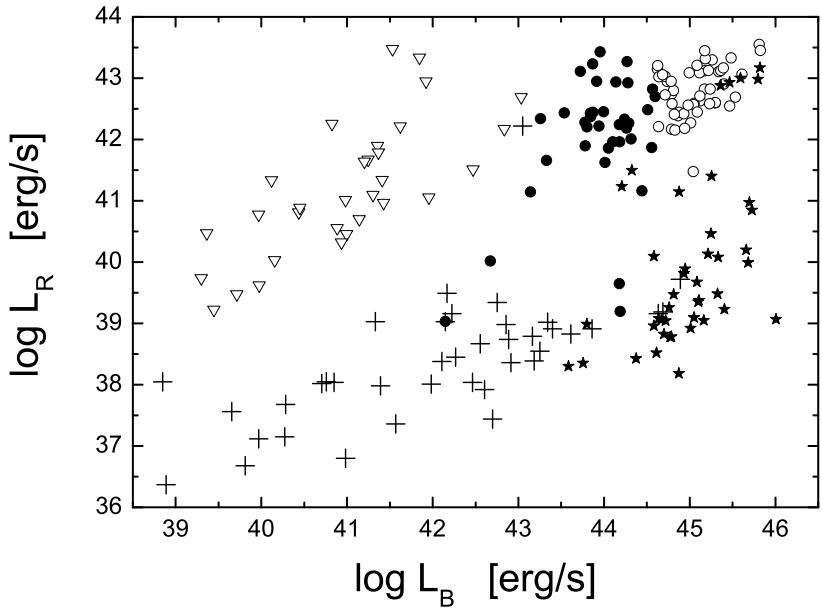}}
 \caption{Total $5$ GHz luminosity vs. $B$-band nuclear luminosity.
BLRGs are marked by filled circles, radio-loud quasars by open
circles, Seyfert galaxies and LINERS by crosses, FR I radio galaxies
by open triangles, and PG Quasars by filled stars. From~\cite{ssl}}
 \label{fig:sikora}
\end{figure}
The jet launching mechanism is unknown. This is rather embarrassing
and some well-intentioned authors prefer to write that it is the
\textsl{details} of this mechanism that are unknown, but this is a
rather huge understatement. In most models of jet launching the
accretion-flow anchored magnetic field plays a crucial role.

Not all accreting compact objects show jet activity. In fact most
quasars are radio-quiet (i.e. do not eject powerful jets). It is not
clear if the distribution of radio-loudness is bi-modal, as had been
believed for a long time (see \cite{ssl} and references therein),
but there are certainly quasars with very similar optical properties
whose radio luminosity differs by several orders of magnitude (see
Fig. \ref{fig:sikora}). Also compact binaries do not always show
jets. In their case jet activity seems to correlate with accretion
luminosity and with the spectral states of the accretion
flow\cite{fender}. The advantage of LMXBs is their short timescales
of variations, which makes it possible to keep track of their
various states in real time. The drawback is their scarcity (e.g.
the correlation found in \cite{gfp} is based on observations of two
systems). The advantage of quasars and Active Galactic Nuclei (AGN)
is their large numbers and variety of types, but their relevant
timescales are much too long to be of any use in direct
observations. Therefore LMXBs in which one can observe jets
appearing and vanishing exert an irresistible attraction and tempt
one to sometimes hasty generalizations. For example, Gallo, Fender,
\& Pooley~\cite{gfp} found that radio luminosities $L_X$ of
black-hole binaries at low accretion rates correlate with X-ray
luminosities, $L_X$. In two systems, GX 339-4 and V404 Cyg, the
relation $L_R \propto L_X^{0.7}$ holds for more than three orders of
magnitude in $L_X$ with the same normalization within a factor of
2.5. This discovery led to speculations that jet activity in LMXBs
is entirely determined by accretion. At least at low accretion
rates, or more exactly during so called hard/low states. At higher
rates, in the so called high/soft state, jet activity is suppressed.
This conclusion was extended to AGN and quasars whose luminosities
were supposed to follow the same type of single relation. However,
as found recently by Sikora, Stawarz \& Lasota~\cite{ssl} in the
case of AGN there are \textsl{two} such relations (Fig.
\ref{fig:sikora}). They have similar slopes and at higher
luminosities one notices in both classes of objects signs of
saturation and intermittency.

The two AGN sequences have rather interesting properties. The
plotted sample consists of radio-loud broad-line AGNs (Broad-Line
Radio Galaxies and radio-loud quasars);  Seyfert galaxies and
LINERS; FR I radio galaxies; and optically selected quasars, and
contains active nuclei hosted by both elliptical and disc galaxies.
The sample \textsl{does not} include blazars, i.e., OVV-quasars,
HP-quasars, and BL Lac objects, because their observed emission is
significantly Doppler boosted, a property too often forgotten in
attempts to find general correlations extending from binaries to
quasars. The two sequences are separated by $\sim 3$ orders of
magnitude in radio luminosity. Manifestly in AGN an additional
parameter is at work in jet production.

All AGN hosted by disc-galaxy (i.e. Seyfert galaxies and LINERS),
including those that according to some criteria are ``radio-loud",
are found only on the lower (``radio-quiet") branch.  AGN hosted by
elliptic-galaxy (quasars and radio- galaxies), however, are found on
both sequences, i.e. they can be radio-loud or radio-quiet. It had
been believed that all AGNs in giant elliptical galaxies are
radio-loud and only recently, using the HST, luminous, radio-quiet
quasars have been found to be hosted by giant elliptical galaxies.
Clearly all this supports the idea that an extra parameter must play
a role in explaining why the upper, radio-loud sequence is reachable
only by AGNs hosted by early type galaxies.

\subsection{Spin paradigm revisited}

There is no much choice in additional parameters. The most natural
one is obviously the black-hole's spin. The idea of the ``spin
paradigm" (like all ideas in about black-hole accretion and
ejection) is not new~\cite{blandford} and it went through various
phases of popularity. Wilson \& Colbert~\cite{wc} assumed that
black-hole spin evolution is determined by mergers and argued that
this produces a broad, `heavy-bottom' distribution of the spin,
consistent with a distribution of radio-loudness. On the other hand
Moderski \& Sikora~(\cite{ms}, see also \cite{msl}) showed that
spin-up by accretion (neglected in \cite{wc}) is so efficient, that
to match the distribution of radio-loudness to observed spin
distribution (i.e. to obtain a large proportion of radio-quiet
quasars corresponding to low spins), one has to assume that
accretion events take the form of both co-rotating and
counter-rotating discs. Both these conclusions have been strongly
challenged, however. On the one hand it was found~\cite{hbspin}
that, unless the merging binary's more massive member already spins
rapidly and the merger with the smaller hole is consistently near
prograde, or if the binary's mass ratio approaches unity, mergers
typically spin black holes down. On the other, Volonteri et
al.\cite{volonteri} argued that the angular-momentum coupling
between black holes and accretion discs is so strong, that the
innermost parts of a disc are always forced to co-rotate with the
black hole, and therefore that all AGN black-holes should have large
spins. In a sense the situation is now inverted: it is apparently
impossible to spin up black-holes by mergers but, nonetheless,
nothing can stop them to spin very rapidly. If true, the
implications are important: if all AGN black holes have very large
spins, then their masses did not grow through mergers and, maybe
less importantly, jet production has nothing to do with the spin of
the central black hole. The last point is of course strengthened by
the presence of relativistic jets in neutron-star LMXBs.

However, the fact that radio-loudness of galactic centers depends on
the host-galaxy morphology makes it worth trying to revive the spin
paradigm. In its framework the much lower radio-loudness of AGNs
hosted by disc galaxies implies (very) low spins of their central
black-holes. One has therefore to elucidate how in an
accretion-dominated evolution, black holes in disc galaxies are
protected against spinning up, whereas those in elliptical galaxies
are not. The physics of the angular-momentum coupling between
black-holes and accretion discs is notoriously complex (see e.g.
\cite{klop} and references therein) but one can expect that if
nuclei of disc galaxies evolve through a sequence of randomly
oriented \textsl{short} accretion events will result in small values
of black-hole spins\cite{ms}\cite{msl}. The required shortness of
the accretion events can be expressed (Sikora et al. in preparation)
in terms of the accreted-mass increments:
\begin{equation}
\frac{m_{\rm acc}}{M_{\rm BH}} << a\sqrt{\frac{R_S}{R_w}},
\label{delta_m}
\end{equation}
where $a=Jc/GM_{\rm BH}^2$ is the dimensionless black-hole spin and
$R_w$ the warp radius\cite{natarprin}.

However, if in contrast to disc galaxies, elliptical galaxies
undergo at least one major merger during their evolution (see, e.g.
ref.~\cite{hbhe06}) the following mass-accretion event is too
massive to satisfy the condition Eq. (\ref{delta_m}). Then
regardless of the initial relative disc-hole orientation the
alignment processes\cite{natarprin} will finally produce co-rotating
configurations and, provided that $m_{\rm acc}/M_{\rm BH}\sim 1$,
this will result in $a\sim 1$.

In black-hole LMXBs the situation is simpler: since to reach the
maximum spin a black hole has at least to double its mass (see
below), black-hole spins in low-mass binaries do not evolve during
the lifetime of the systems. Therefore, if black-holes in LMXBs are
born with roughly the same spin (or at least with no bimodal
distribution), one should not expect in this case the presence of
two radio-loudness sequences as observed~\cite{gfp}.

\subsection{Jets from neutron-star X-ray binaries}

Relativistic jets are also observed in NS LMXBs~\cite{fender}.
Neutron stars have no ergospheres so the ejection mechanism cannot
be the exactly the same as in BH systems. However, the condition for
launching a Poynting-flux dominated outflow, which later becomes
converted to the matter dominated relativistic jet, is to supply a
high magnetic--to--rest-mass energy ratio ($\gg 1$) at the base of
the outflow. This is obviously satisfied in the case of the magnetic
field anchored on a (rotating) neutron star (see \cite{ssl} for more
details).

However, it is rather amusing to notice that when a jet-like
structure is observed in an X-ray binary it is immediately assumed
that the compact component is a black-hole. This allows to call the
system a "micro-quasar" which sounds good in a press release. Of
course this is possible only if there is no direct evidence that the
compact body is a neutron star (incidentally there is no
\textsl{direct} evidence that the compact body in Sco X-1 and
Cir~X-1, the two ``radio-loud" neutron-star binaries is indeed a
neutron star but somehow nobody, as far as I know, tried to claim
that they could be micro-quasars.) So when very high energy
$\gamma$-rays were observed with HESS \cite{ls50} from the X-ray
binary LS 5039 which exhibits jet-like structure, it seemed
inevitable that this was due to particle acceleration in a
microquasar jet. However, as showed by Dubus\cite{gd50} the compact
object in this \textsl{high-mass binary} is a young pulsar. In a
similar system the object PSR B1259-63 is a pulsar and a third
system LSI +61.303 also belongs to this category of gamma-ray
binaries. A recent criticism\cite{mirabel} of the model presented by
Dubus is not very thoughtful. It is true that the pulsar model
underestimates the $\gamma$-ray fluxes but this can be explained by
the simplifying assumptions (isotropy, no pair cascades) and in any
case the micro-quasar model is not doing better. The argument that
``jets" in LS 5039 seem to have relativistic bulk motions as in
micro-quasars is irrelevant because it assumes that these jet-like
features \textsl{are} jets and their speeds then deduced from their
asymmetry. In the pulsar model, however, these features are not jets
but radio-tails that just mimic (micro-quasar) jets. Finally the
absence of major radio outbursts in LS 5039 is not an argument
against the pulsar model but simply a consequence of the fact that
in this system the stellar companion is an $O$ star so that the
pulsar have no circumstellar disc to plunge through, whereas its
cousins in PSR B1259-63 and LSI +61.303 enjoy the presence of $Be$
companions and can therefore produce radio splashes.

\subsection{Black hole spin-up}

It is (too) often forgotten that although to spin a black-hole up to
the maximum rotation-rate through accretion from a Keplerian disc is
in principle easy, it is \textsl{impossible} in LMXBs. The reason is
simple and has been known for 36 years: to spin-up to maximum rate a
black-hole must accrete more than twice its mass \cite{bardeen}.
Since apparently this is not universally known and also the
difference in this respect between black holes and neutron-stars is
not clear to everybody, it is worth showing it once more. Bardeen's
solution can been found in \cite{volonteri} - e.g. an initially
non-rotating black hole gets spun up to the maximum rate after
accreting $\sqrt{6}$ of the initial mass $M_0$. It is indeed a
``modest" \cite{volonteri} amount by extragalactic standards but it
is at least several times more than the mass of the black-hole
companion in LMXBs. Here some simple reasoning will show the
difference between spinning up a neutron star and a black hole.

In the case of a neutron star, assuming the disc extends to the
star's surface, the mass $\Delta M $ accreted during the spin-up to
an angular frequency $\Omega_{*}$ can be expressed as
\begin{equation}
\Delta M \approx \frac{I_{*}
\Omega_{*}}{\left(GM_{*}R_{*}\right)^{1/2}}, \label{nsspinup}
\end{equation}
where $I_{*}$, $M_{*}$ and $R_{*}$ are the neutron star's moment of
inertia, mass and radius respectively. It is therefore enough to
accrete $0.1 \msol$ to spin up a $1.4 \msol$ neutron star to
millisecond periods. More generally, to spin up a neutron star to
break-up one needs to accrete:
\begin{equation}
\Delta M \approx \alpha(x) M_{*}
\end{equation}
where I have used $I_{*}\approx \alpha(x) M_{*}R_{*}^2$
~\cite{bhae}. $x=(M{*}/\msol) (km/R_{*})< 0.24$\cite{maxz} is the
compactness parameter for the most compact neutron star
$\alpha(x)\lta 0.489$~\cite{bhae}.

One can use Eq. (\ref{nsspinup}) also for a black hole but one
should now use the formula for the angular velocity at the horizon
\begin{equation}
\Omega_{\rm H}= \frac{a}{2Mr_+}, \label{omegah}
\end{equation}
where $a=J/M$ and $r_+ = M + \sqrt{M^2-a^2}$ is the black hole
(outer horizon) radius. For a maximally rotating black hole $a=M$,
$r_+=M$, therefore writing $I=r_+^2= 1/M$ one obtains
\begin{equation}
\Delta M = {M}_1 - M_0 \approx \frac{1}{2}M_1
\label{dm}
\end{equation}
It might seem surprising that one obtains a sensible result in such
a basically newtonian way, but the $\Delta M$ from Eq. \ref{dm}
corresponds to the inequality expressing the black-hole surface area
theorem:
\begin{equation}
\Delta M > \Omega_{\rm H} \Delta J
\end{equation}
A more refined treatment of spin-up of black holes in LMXBs can be
found in ref.~\cite{kk}. (A black hole initially counterotating with
respect to the accretion disc must triple its mass to achieve
maximum spin \cite{klop}.)

Therefore black holes in LMXBs keep their inborn spin. Observations
seem to suggest that it is not very close to maximal ($a/M \sim 0.1
- 0.8$~\cite{spin-bh}~\cite{done} but one should remember that such
conclusions are strongly model-dependent. Simulations suggest
formation of stellar-mass black holes with $j=J/M\sim
0.7$~\cite{gammie}.



\section*{Acknowledgements}
I am grateful to Mike Garcia for sharing with me his
minimum-luminosity database and to Didier Barret, Guillaume Dubus,
Jean-Marie Hameury and Marek Sikora for enlightening discussions.
Albert Lazzarini is thanked for helpful criticism. I am grateful to
George Chapline for his insights about Dark Energy Stars. Marek
Abramowicz provided constant encouragement and intellectual support.
I was financially supported in part by the CNES and the GDR--PCHE of
the CNRS.

\end{document}